\documentclass[british]{llncs}
\usepackage[T1]{fontenc}
\usepackage[latin9]{inputenc}
\usepackage{listings}
\usepackage{array}
\usepackage{verbatim}
\usepackage{graphicx}
\usepackage{setspace}

\makeatletter

\providecommand{\tabularnewline}{\\}

\@ifundefined{definecolor}
 {\usepackage{color}}{}
\usepackage{ifpdf} 
\ifpdf 

 \IfFileExists{lmodern.sty}{\usepackage{lmodern}}{}

\fi 

\usepackage{fancyvrb}

\makeatother

\usepackage{babel}

\begin{document}

\title{A Test Automation Framework for Mercury}

\author{Peter Biener\and Fran\c{c}ois Degrave%
\thanks{Supported by a grant FRIA - Belgium.%
} \and Wim Vanhoof}

\institute{Faculty of Computer Science, University of Namur, Belgium}
\maketitle
\begin{abstract}
This paper presents a test automation framework for Mercury programs.
We developed a method that generates runnable Mercury code from a
formalized test suite, and which code provides a report on execution
about the success of test cases. We also developed a coverage tool
for the framework, which identifies and provide a visualization of the reached parts of the program when
executing a given test suite.
\end{abstract}

\section{Introduction}

Testing is today's most commonly used method for finding defects and
increase quality of software. It has been in focus for many years,
since correcting defects is often the most substantial part of the
software development budget \cite{book:glass}. When developing business
critical systems, it is particularly important to detect bugs in time.
Additionally, in this sector the release process is much stricter
than usually, so detecting a bug only at a later phase of the release
process can postpone the release date by months or years. %


{}While the fact that the system under test successfully passes a large
number of tests does not prove correctness of the software, it nevertheless
increases confidence in its correctness and reliability \cite{book:kaner}.

%

In testing terminology, a \emph{test case} for a software component refers to the combination
of a single test input and the expected result. A \emph{test
suite} refers to a collection of individual test cases. Evaluating
a test suite is a process that can be automated by using a tool that
runs the software component that is being tested once for each test
input, compares the actual result with the expected result and reports
those test cases that failed during the test; the most well-known
example of such a tool is JUnit for Java \cite{12hunt}. Such an automation
framework allows the repeated evaluation of a test suite, for example
to perform so-called regression testing -- in order to detect errors
that are introduced by changing code that was previously working fine.

%

There exist several test frameworks for declarative programming languages.
For example, Prolog Unit Tests \cite{SWI:module_test} -- an integrated
test framework for SWI-Prolog --, the basic \texttt{test\_util} library
for ECLiPSe Prolog, and HUnit \cite{Haskell:HUnit} for Haskell. The
target of this work is to develop a test automation framework for
the Mercury language, which -- to the best of our knowledge -- has
no such tool available yet. In our work we follow some principles
from the mentioned tools, though we cannot port any of those tools
directly to Mercury because of language specialities. For example, the strict type- and mode-checking
mechanisms make it difficult to adopt most of the methods used in
Prolog, even if we can of course can re-use some of the ideas in the design phase.%


The main goal of an integrated test framework is to interpret a previously
formalized test suite, execute the independent test cases, and finally
produce a report about which test cases failed and why. 
However, a test framework can also have additional
features, which are not necessarily needed for the testing itself;
a module could for example interact with an integrated debugger in
order to try to identify the code fragment that caused the failure
of a given test case \cite{55841}.
It can also provide a coverage tool, i.e. a tool which
is able to produce a measure describing the degree to which the source
code of the program has been exercised during the execution a given
test suite.

This measure is performed with respect to one or more \emph{coverage
criteria} \cite{zhu:coverage}. Among the most used coverage criteria,
the \emph{procedure} coverage criterion aims at verifying if every
procedure is called, while the \emph{entry/exit} coverage criterion
is similar to the latter but takes into account the success or failure
of a procedure's execution -- which makes it particularly appropriate
when testing logic programs%
. The \emph{statement} coverage \cite{winko:Myers2004} criterion
is relatively simple as it measures if every statement or instruction
is reached during execution, whereas \emph{branch} (or block) coverage
criterion requires every condition of if-then-else statements to be
evaluated to every possible value (usually true and false) (a variant
of this criteria is used in the Emma coverage tool for Java \cite{Emma:reference_manual}).
The most general coverage criterion is the \emph{path} coverage criterion,
which requires every possible route through the code to be executed.
Note that full path coverage is in general impossible; indeed, on
the one hand loop constructs can result in an infinite number of paths,
on the other hand the program under test can contain unreachable code
fragments.

\paragraph*{}
In what follows, we first introduce some background knowledge about the Mercury language (Section~\ref{prelim}), then we present the unit testing tool and the interesting characteristics of its implementation (Section~\ref{unittest}). In Section~\ref{sec:covtool} we present the details of the coverage tool, then we show and discuss the results of the evaluation of the prototype (Section~\ref{eval}).

\section{Preliminaries}\label{prelim}

Mercury is a statically typed logic programming language \cite{mercury:jlp}.
Its type system is based on polymorphic many-sorted logic and is essentially
equivalent to the Mycroft-O'Keefe type system \cite{mycroft:typesprolog}.
A type definition defines a possibly polymorphic type by giving the
set of function symbols to which variables of that type may be bound
as well as the type of the arguments of those functors \cite{mercury:jlp}.
Take for example the definition of the well known polymorphic type
$list(T)$:

\begin{lstlisting}[basicstyle={\ttfamily},breaklines=true,tabsize=4]
:- type list(T) ---> [] ; [T|list(T)].
\end{lstlisting}

\noindent According to this definition, if $T$ is a type representing
a given set of terms, values of type $list(T)$ are either the empty
list \texttt{{[}{]}} or a term $[t_{1}|t_{2}]$ where $t_{1}$ is
of type $T$ and $t_{2}$ of type $list(T)$.

In addition to these so-called \emph{algebraic types}, Mercury defines
a number of primitive types that are builtin in the system. Among
these are the \emph{numeric types} \texttt{int} (integers) and \texttt{float}
(floating point numbers). Mercury programs are statically typed: the
programmer declares the type of every argument of every predicate
and from this information the compiler infers the type of every local
variable and verifies that the program is well-typed.

In addition, the Mercury \emph{mode system} describes how the instantiation
of a variable changes over the execution of a goal. Each predicate
argument is classified as either input (ground term before and after
a call) or output (free variable at the time of the call that will
be instantiated to a ground term). A predicate may have more than
one mode, each mode representing a particular usage of the predicate.
Each such mode is called a \emph{procedure} in Mercury terminology.
Each procedure has a declared (or inferred) \emph{determinism} stating
the number of solutions it can generate and whether it can fail. Determinisms
supported by Mercury include \texttt{det} (a call to the procedure
will succeed exactly once), \texttt{semidet} (a call will either succeed
once or fail), \texttt{multi} (a call will generate one or more solutions),
and \texttt{nondet} (a call can either fail or generate one or more
solutions)%
\footnote{There exist other modes and determinisms but they are outside the
scope of this paper; we refer to \cite{mercury:jlp} for details%
}. Let us consider for example the definition of the well-known \texttt{append/3}
and \texttt{member/2} predicates. We provide two mode declarations
for each predicate, reflecting their most common usages:

\begin{lstlisting}[basicstyle={\small\ttfamily},breaklines=true,tabsize=4]
:- pred append (list(T), list(T), list(T)).
:- mode append(in, in, out) is det.
:- mode append(out, out, in) is multi.

append([], Y, Y).
append([E|Es], Y, [E|Zs]):- append(Es, Y, Zs).


:- pred member(T, list(T)).
:- mode member(in, in) is semidet.
:- mode member(out, in) is nondet.

member(X, [X|_]).
member(X, [Y|T]) :- not (X=Y), member(X, T).
\end{lstlisting}

For \texttt{append/3}, either the first two arguments are input and
the third one is output in which case the call is deterministic (it
will succeed exactly once), or the third argument is input and the
first two are output in which case the call may generate multiple
solutions. Note that no call to \texttt{append/3} in either of these
modes can fail. For \texttt{member/2}, either both arguments are input
and the call will either succeed once or fail, or only the second
argument is input, in which case the call can fail, or generate one
or more solutions. Note that unlike Prolog, Mercury doesn't handle partially instantiated data structures.

\section{Unit testing tool for Mercury}\label{unittest}

The goal of this work is to create a
framework for Mercury that lets the user define test cases through
a simple language, and from that point on, automatically performs the whole testing process. In our implementation, the framework has two independent
modules: one is responsible for executing a test suite, the other
is a coverage tool, which is presented later in this paper. The two
modules are loosely connected, each one being usable without the other.

The generation process is completely independent
from the tested code: one can write test cases without any knowledge
of the source code, so the tool is even usable for test-driven development.

\begin{figure}
\begin{centering}
\includegraphics[bb=10bp 265bp 665bp 525bp,clip,scale=0.5]{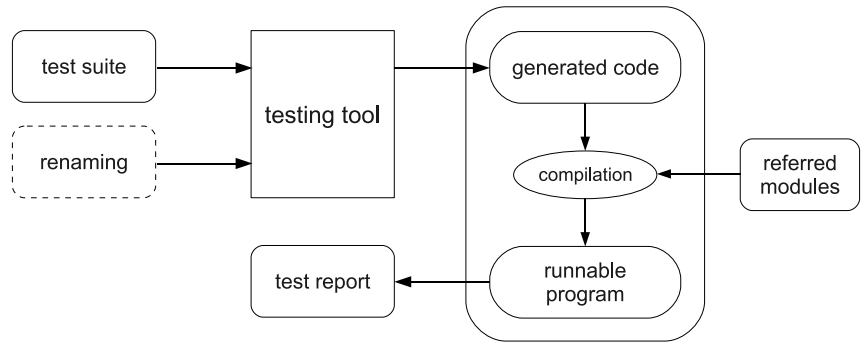}
\par\end{centering}

\caption{\label{fig:Testing-framework}Testing framework}

\end{figure}

A schematic diagram of the testing process is shown on Figure~\ref{fig:Testing-framework}.
The test cases are contained in the test suite file. 
The other (optional) input of the tool is a renaming information file, which is generated
by the coverage tool. Its usage is explained in Section~\ref{sec:covtool}.
In general, a formal test case representation consists of a unique
test case, i.e. the combination of a code fragment to be executed together with one or more assertions
on the expected results.
In our implementation, a test case is a triple, written as $\mbox{test}(t,c,a)$,
where $t$ is the name of the test case (a Mercury string), $c$ is
a Mercury code fragment represented as a list of atoms, and $a$ is a
list of assertions. An assertion can be either a condition on the
variables of $c$, or a specification of the expected behaviour of
the execution. For the latter, the test framework provides three options: \texttt{\textbf{succeed}},
\texttt{\textbf{fail}} or \texttt{\textbf{exception}}. 

Let's examine a simple example of the syntax of a test case:

\begin{example}\label{ex:testcase}\ \\
\vspace*{-0.5 cm}
\begin{lstlisting}[basicstyle={\ttfamily},breaklines=true,tabsize=4]
test(t1, [reverse([1,2],L)], [true(L=[2,1])]).
\end{lstlisting}

\noindent \texttt{t1} is the name that will be used to refer to the
test case in the report generated by the testing tool. The code fragment
to test contains only one goal (a call to the list reverse predicate),
while the only assertion is a condition verifying whether the value
computed for \texttt{L} is indeed the result of reversing the list
\texttt{{[}1,2{]}}. 
\end{example}


%

\subsection{Determinism}

If only the features mentioned above are used then execution of the
test code is limited to the first solution, even if the predicate under concern has possibly multiple solutions. In the latter cases, all the solutions but the first one are dropped. Nevertheless, more
extensive examination of \texttt{multi} or \texttt{nondet} predicates
is also possible. In order to achieve this, the following conditions can be used in
the assertions part:
\begin{description}
\item [{true(C)}] Simple execution of C. Only a single instantiation of
the assertion C is verified, namely the one obtained from the first
answer returned by the tested code in the test case.
\item [{some\_true(C)}] The given condition holds for at least one solution
of the tested code.
\item [{all\_true(C)}] The given condition holds for all solutions.
\item [{true(N,~C)}] The given condition holds for the Nth solution%
\footnote{This is definitely not a pure declarative condition, since the result
may depend on the order of results, and thus on the Mercury implementation.%
}.
\item [{solutions\_cardinality(N)}] N is equal to the number of solutions.
N can be either a variable or a constant.
\item [{type(V,~T)}] User defined type information of a variable.
\item [{limit(N)}] Limits the execution to N solutions. This can be useful
when testing predicates with a large number
of solutions.
\end{description}
Example~\ref{ex:testcases2} shows the usage of some of these conditions. 

\begin{example}\label{ex:testcases2}\ \\
\vspace*{-0.5 cm}
\begin{lstlisting}[basicstyle={\ttfamily},breaklines=true,tabsize=4]
test(t2, [member(X,[1,3,4,2])], 
	[limit(2),some_true(X>1)]).
test(t3, [member(X,[1,2,3,4])], 
	[solutions_cardinality(N),true(N>3),
	all_true(X<5)]).
\end{lstlisting}

The
semantics of the assertions in \texttt{t2} is \textquotedblleft{}there
is a solution among the first 2 that is bigger than 1\textquotedblright{},
while the meaning of the assertions in \texttt{t3} is \textquotedblleft{}there
are at least 3 solutions, and all of the solutions are less than 5\textquotedblright{}.

\end{example}

Usage of input/output operations is not permitted in the assertions part, but they are allowed in the tested code fragment under the condition that
 all the goals are deterministic. In practice it means that there are two different
execution modes of the test tool: {}``multi'' and {}``IO''. With
{}``multi'', testing multi-solution predicates is possible, but
usage of IO is not. With {}``IO'', it is the other way round. This
execution mode of the tool can be chosen by a command line option.

\subsection{Implementation details}

Figure~\ref{fig:Generated-code-det} shows the generated code for
the test case \texttt{t1} defined in Example~\ref{ex:testcase}. Since the expected behaviour
is specified as success, if the \texttt{reverse/2} predicate fails, the result of the test case will be {}``failed
because of failure (instead of success)''.

\begin{figure}[h]
{\scriptsize 
\begin{lstlisting}[basicstyle={\small\ttfamily},breaklines=true]
...
testcase(t1, Result) :-
  (if
    reverse([1,2], L)
  then
    (if 
      L = [2,1]
    then
      Result = succeeded
    else
      Result = condition_failed
    )
  else
    Result = failed(failure)
  ).
\end{lstlisting}
}\caption{\label{fig:Generated-code-det}Generated code (det)}

\end{figure}

Testing multi-solution predicates needs some considerations. Our implementation
uses Mercury's \texttt{solutions} library for handling predicates
that can succeed more than once. However, this library has an important
restriction, namely that the given predicate can have only one output
argument. An easy solution to this problem is to wrap all the output
variables into a compound term, then
unwrap the variables after execution of the code and then perform the checks of the assertions part. Unfortunately,
for the generation of this compound type declaration, the type of
each output variable should be known. The need of type analysis could strongly limit
the usability of the tool since all the
sources of used modules should be known in that case. This is usually not feasible, especially
in case of built-in modules. The workaround we developed is to use the type analysis facility of the compiler itself. It is
possible with the \texttt{univ} library, which allows to wrap any Mercury type
into a universal type. For the unwrap operation, the compiler must
know the type of the wrapped object. Usually, it can be inferred from
the assertions, but if not, we have to give the type manually, as
a help to the compiler. Example~\ref{ex:testcases3} shows the usage of this feature.

\begin{example}\label{ex:testcases3}\ \\
\vspace*{-0.5 cm}
\begin{lstlisting}[basicstyle={\ttfamily},breaklines=true,tabsize=4]
test(t4, [append(L1,L2,[1,2,3])], [type(L2,list(int)), 
	some_true((L1=[1,2],length(L2,1)))]).
test(t5, [append(L1,L2,[1,2,3])], 
	[some_true((L1=[1,2],L2=[3]))]).
\end{lstlisting}

\noindent The tested code fragment is the same in both test cases:
the \texttt{(out,out,in)} mode of \texttt{append/3}. In \texttt{t4}, the compiler
can infer the type of \texttt{L1}, but the type of \texttt{L2} must
be given explicitly. In the other test case, the compiler doesn't
need any complementary information. Notice that if there is no
more than one common variable between the two code parts, then no
wrapping is used, and thus no type information needs to be provided.
\end{example}

Generation of code for the \texttt{some\_true/1}, \texttt{all\_true/1}
and \texttt{true/2} conditions is based on the same principle. Unwrap
instructions of output variables are appended before the given condition
if necessary, then this code fragment -- a meta-predicate -- is called
in an appropriate way. For example in the case of \texttt{true/2}, after selecting
the required solution from the list, the constructed meta-predicate
is called simply using the \texttt{call/2} predicate. 
The optional solution
number limitation is implemented with the help of \texttt{do\_while}
predicate in the \texttt{solutions} library.

Figure~\ref{fig:Generated-code-nondet} shows the generated code
for the test case \texttt{t4}, where we can see the declaration for
the generated type. The two lines just after the call to \texttt{append/3}
wrap the output variables into a single compound term. The reverse
operation is performed by the two lines just before the assertions. The combination of the latter together with the assertions themselves constitute the body of a meta-predicate.
This meta-predicate must succeed for at least one solution for
the test case to be considered as successful.

\begin{figure}[h]
{\scriptsize 
\begin{lstlisting}[basicstyle={\small\ttfamily},breaklines=true]
...
:- type t4_type ---> t4_t(univ, list(int)).
testcase(t4, Result) :-
  solutions( ((pred (IF1 :: out)) is nondet :-
      append(L1, L2, [1,2,3]),
      type_to_univ(L1, L1_U),
      IF1 = t4_t(L1_U, L2)
    ), Vs),
  (if
    some_true( ((pred (IF2 :: in)) is nondet :- 
        IF2 = t4_t(L1_U, L2),
        det_univ_to_type(L1_U, L1),
        L1 = [1,2], length(L2,1)
      ), Vs)
  then
    Result = succeeded
  else
    Result = condition_failed
  ).
\end{lstlisting}
}\caption{\label{fig:Generated-code-nondet}Generated code (nondet)}

\end{figure}

\subsection{Handling exceptions}

By default, every exception thrown by either the tested code or some
of the assertions is caught by the framework. This is necessary,
otherwise it could interrupt the execution of a large test suite with
possibly a very large number of test cases. In special cases, the expected
result can be exception too, which is a feature the framework can
handle with a small restriction: exceptions cannot be
distinguished by their origin, one thrown by an assertion is handled
in the same way as if it had been thrown by the tested code. Currently
it is impossible to make assertions on the exception itself, the only
thing that can be declared in the assertion part is {}``the code must
throw an exception''.

Nevertheless, it can happen that exceptions need to be left uncaught,
especially when the user wants to know the exact source of an exception,
usually to know where to find a given bug. 
If the
exception is caught, the result will only be {}``the test case threw
an exception'', but the real cause remains hidden. To help to identify
these problems, exception handling mechanism can be entirely switched
off by a command line switch, so in that {}``debug'' mode, the details
of the problem become observable.

\section{Coverage tool}\label{sec:covtool}

This tool is a complementary module for the base framework that helps
to detect parts of the tested code which are uncovered by a given
set of test cases. Logic programming languages have a few peculiarities
that must be taken into account when constructing a coverage tool. The most important of these is nondeterminism, namely that
statements can fail and/or have multiple solutions. Most coverage
tools for declarative languages transform the original program to
an instrumented code and place some kind of execution counters before
and after calls. This enables tracing calls to and exits from
procedures. The counters are usually stored in a non-declarative way,
like in the Haskell Program Coverage tool \cite{DBLP:conf/haskell/GillR07},
which records every increment into a file. This is unavoidable in
case of logic programming languages, since after a backtracking, all
changes made on pure declarative variables would be revoked. The same
principle is used in the \texttt{coverage} library for ECLiPSe Prolog,
the output of which is a simple HTML page, where the counter values
are shown between the goals under concern.

Our tool follows the same principles as these tools, but needs
to deal with some particularities of the Mercury language. The most notable one is the mode-reordering mechanism of the Mercury compiler,
as it strongly affects coverage. Another issue that is worth mentioning
is the way that the Mercury compiler treats switches. 
In the next section, we expose these different issues and present the solutions we developed.

\subsection{Implementation}

In the remaining, we assume that the programs are well-moded; this condition is checked during compilation by the Mercury compiler. The latter also re-orders the goals in such a way that they are executed from left to right.
Since different modes
usually imply different orders, multi-moded predicates must be transformed
into several different procedures.

\begin{figure}[h]
\begin{centering}
\includegraphics[bb=15bp 205bp 578bp 770bp,clip,scale=0.5]{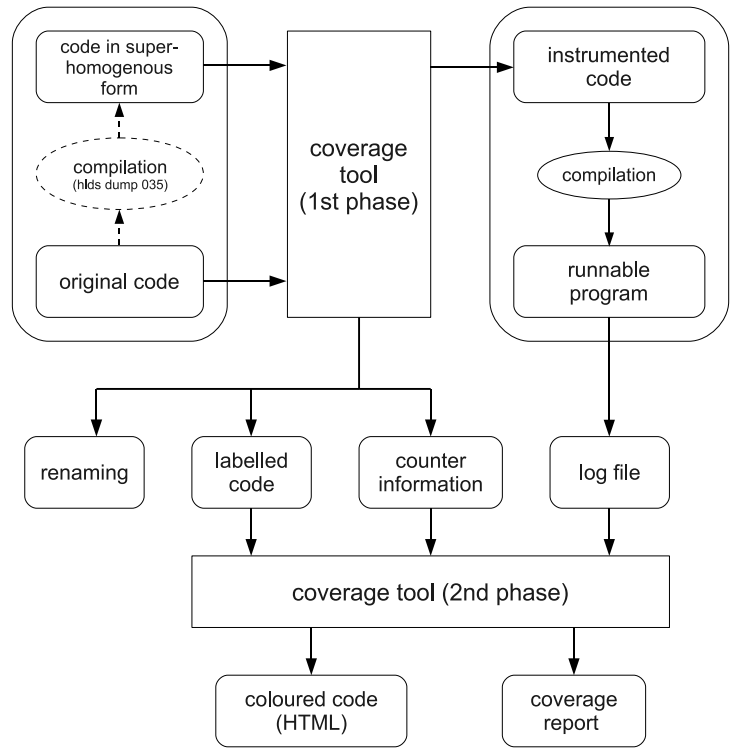}
\par\end{centering}

\caption{\label{fig:Coverage-tool}Coverage tool}

\end{figure}

Our implementation transforms the examined code into an instrumented
one, compiles it and executes it in order to log execution
information; the process of coverage measuring is shown in
Figure~\ref{fig:Coverage-tool}. The base idea of the transformation
is to add counters in the code, implemented by logging calls that write unique identifiers into a log file. The counters are
placed with respect to the labelled superhomogenous
form syntax defined in \cite{DBLP:conf/lopstr/DegraveSV08}, with minor simplifications:

\medskip{}

\begin{definition}
\noindent \textit{The syntax of a program in labelled superhomogenous
form is defined as follows:}

{\small \[
\begin{array}{llcl}
LProc &  & ::= & p(X_{1},\ldots,X_{k})\texttt{ :- }LConj.\\
LConj & C & ::= & _{l}G_{l'}\ |\ _{l}G\texttt{,}C\\
LDisj & D & ::= & C\texttt{;}C''\ |\ D\texttt{;}C\\
LGoal & G & ::= & A\ |\ D\ |\ \texttt{not}(C)\ |\ \texttt{if}\ C\ \texttt{then}\ C'\ \texttt{else}\ C''\\
Atom & A & ::= & X=Y\ |\ X=f(Y_{1},...,Y_{n})\ |\ p(X_{1},...,X_{n})\end{array}\]
 }{\small \par}
\end{definition}

A counter is assigned to each label $l$, denoted by $\textrm{counter}(l)$.
Basically this means that counters are inserted into every possible
place between goals.

The first step of the transformation process is to get
the code to be in superhomogeneous form. A part of this process can be achieved by the compiler (goals reordering, duplication of predicates with multiple modes); however all the multi-moded predicates need to be renamed, in such a way that every procedure is associated to
a unique name. Every call to the procedures must therefore be renamed consequently; this
can be done thanks to a simplified mode analysis, following the instantiations
of variables throughout the code. The new name assignments are saved
into a file, in order to provide names mapping information to the user at the end of the process.

\subsection{Switches vs disjunctions}

The second step of the transformation is the addition of logger calls between goals of the code; these calls reify incrementing operations on counters: $\textrm{log}(l)\equiv\left\{ \textrm{counter}(l):=\textrm{counter}(l)+1\right\} $.
Unfortunately, this step can render the program not compilable if it
contains \emph{switches}. A switch is a special disjunction -- with nothing visually distinguishing it from a ``regular'' disjunction --, in which {}``each
disjunct has near its start a unification that tests the same bound
variable against a different function symbol'' \cite{Mercury:reference_manual}. In the remaining, we call such unifications the \emph{switch conditions}.
In a single switch, the switch conditions are exclusive from each other;
that allows the compiler to consider the switch as being deterministic or semi-deterministic -- depending on whether every possible
condition value is covered -- whereas regular disjunctions are, in general, non- or multi-deterministic. Switches can be nested into each other
and if they test the same variables, they are likely to be treated
as a single switch. 

\begin{figure}[th]
\begin{tabular}{>{\centering}p{0.4\linewidth}>{\centering}p{0.03\linewidth}|>{\centering}p{0.03\linewidth}>{\centering}p{0.4\linewidth}}
\begin{doublespace}

\begin{lstlisting}[basicstyle={\ttfamily},lineskip={-3pt},tabsize=4]
...,
(
  X = f,
  p(Out)
;
  Y = X,
  (
    Y = g,
    Intermediate = 42
  ;
    Z = Y,
    Z = h(Arg),
    q(Arg, Intermediate)
  ),
  r(Intermediate, Out)
),
...
\end{lstlisting}
\end{doublespace}
 & $\quad$ & $\quad$ & \begin{doublespace}

\begin{lstlisting}[basicstyle={\ttfamily},lineskip={-3pt},tabsize=4]
(
  log("label_1"),
  X = f,
  log("label_2"),
  p(Out),
  log("label_3")
;
  log("label_4"),
  Y = X,
  log("label_5"),
  (
    log("label_6"),
    Y = g,
    log("label_7"),
    Intermediate = 42,
    log("label_8")
  ;
    log("label_9"),
    Z = Y,
    log("label_10"),
    Z = h(Arg),
    log("label_11"),
    q(Arg, Intermediate),
    log("label_12")
  ),
  log("label_13"),
  r(Intermediate, Out),
  log("label_14")
)
\end{lstlisting}
\end{doublespace}
\tabularnewline
\end{tabular}

\caption{\label{fig:Instrumenting-a-switch}Naive instrumentation of a switch}

\end{figure}

The reason switches are considered as particular structures is to allow the compiler to perform a determinism analysis; no regular disjunction is allowed in predicates declared as \emph{det} or \emph{semidet}.
Only unifications can precede switch conditions in the different disjuncts; if not, the compiler is not able to detect the switch conditions, and therefore considers the disjunction under concern as a regular (\emph{nondet} or \emph{multi}) disjunction. 
When logger calls are inserted at the beginning of a disjunct, a switch
will therefore be considered as a regular disjunction. That can cause the compilation to fail if the enclosing predicate
is (semi-)deterministic. The example shown at the left side of Figure~\ref{fig:Instrumenting-a-switch}
is extracted from the Mercury reference manual: it is a switch on X, provided
X is ground. On the right side, there is the ``naively'' instrumented version with
the logger calls; this instrumented code would cause the program to fail at compiling if it is used in a (semi-) deterministic predicate.

\begin{figure}
\begin{centering}
\includegraphics[bb=10bp 463bp 550bp 525bp,clip,scale=0.5]{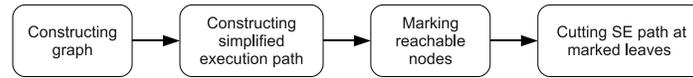} 
\par\end{centering}

\caption{\label{fig:Switch-transformation}Switch transformation}

\end{figure}

The solution we developed is to replace, in a disjunct,
logger calls
before each unification at the beginning of a disjunct by a single logger predicate call \textit{after} the unifications -- just before the first predicate call occurring in the disjunct. This single logger call should update all the counters in order to
reflect success or failure of 
all the preceding disjuncts.
%
%
Unfortunately, this solution has a drawback: if one of the unifications preceding the logger call fails during execution, it is not possible to know which one it was (since no counter was placed between them) and then the coverage information is incomplete. However, we can recover this missing information by performing a small analysis since the values of the variables involved in these unifications are known before entering the disjunction.
%
The algorithm that determines which counter
needs to be updated at which point is presented below (its basic steps
are shown on Figure~\ref{fig:Switch-transformation}).

\begin{figure}
\begin{centering}
\includegraphics[bb=20bp 435bp 500bp 527bp,clip,scale=0.5]{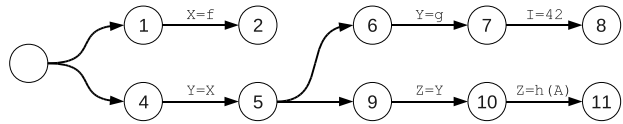} 
\par\end{centering}

\caption{\label{fig:Switch-execution-graph}Switch execution graph}

\end{figure}

We model a switch under the form of a tree, since they can be nested into each
other. Nodes of the tree are the labels of the labelled superhomogenous
form of the program, while its edges are the unifications between the labels. All the
statements after the first predicate call of each disjunct are dropped from the model,
so the leaves of the tree are the labels preceding the first
predicate call in each disjunct. Complex statements,
like conditional structures, etc. are treated as if they were predicate
calls, and are thus also dropped from the model. If there are
only unifications in a disjunct, then the leaf of the corresponding branch is the
last label of the disjunct. We can define a \textit{simplified execution
path} for a switch as a sequence of labels, which is the output of
a depth-first search in the corresponding model graph. The model of
the example of Figure~\ref{fig:Instrumenting-a-switch} is shown in Figure~\ref{fig:Switch-execution-graph},
while the corresponding simplified execution path is (1,2,4,5,6,7,8,9,10,11).

Before executing a switch, all edges of its model graph are examined.
%
If a unification succeeds, then
its successor node is marked, otherwise this part of the tree is left
out from further examination. The first node of each branch is also marked, since they are not assigned to any unification.
After this step, the nodes that
are not marked correspond to program points that are not reached on
the examined program state, and thus no counter update is applied
there. 
The marked nodes are visited using a depth-first search; when a leaf is reached, the sequence of marked nodes encountered on the path up to this leaf is stored and associated to the leaf's label. This process is repeated starting from the next unvisited marked node until no marked node is left unvisited.

Thanks to this method, a simplified execution path is split into sequences, whose last elements
are mapped to program points that are reached on the examined program
state. These are the sequences of labels that must be written out
to the log file when the execution reaches the given label. 

Using the example from Figure~ \ref{fig:Switch-execution-graph} again, and assuming that every unification was successful (which
is of course impossible, since the value of \texttt{X} cannot be \texttt{f},
\texttt{g} and \texttt{h(Arg)} at the same time), the sequences
assigned to leaves would be $(1,2)$ for leaf $2$, $(4,5,6,7,8)$ for leaf $8$, and $(9,10,11)$ for leaf $11$.
If the unification \texttt{X=f} between the first two labels fails,
then the assignments will be (1,4,5,6,7,8) for leaf 8, and (9,10,11) for leaf 11. The original
logging statements can be removed from the code, and only one batch logging
statement is needed at the {}``leaves'' of the switch (in each successful
branch).

The last issue remaining using that technique is due to the last segment of the simplified
execution path when the last node is not marked. It means that the
last branch (or the last few branches) fails, and in this case we
have no information about which unifications were executed. We
decided to log these entries at the same time with the previous batch
logging action, or if there is no successful branch, log
them before the execution of the switch. For example, if the failing
unifications are \texttt{X=f} and \texttt{Z=h(Arg)}, then the only
batch will be (1,4,5,6,7,8,9,10) associated to leaf 8. 

\subsection{Connection with the base framework}

Once the code has been successfully instrumented, using the coverage
tool as a part of the testing framework is easy: in the test suite
file, we simply refer to this transformed code instead of the original
one. For
convenience, it isn't necessary to rename multimode predicates in
test cases, it is possible to pass the renaming information file --
which is generated by the instrumentation step -- to the test automation
module. With this feature, the new predicates will be called on test
execution.

\subsection{Statistics and report}

The direct output of executing instrumented code is a log file, which
contains information about reached program points. However this information is not easily readable; it should appear under a more ``user-friendly'' form in order to be useful. For this purpose there is a second execution phase in our implementation. It makes use of different information sources: the labelled source file (the instrumented code just
before switch transformation -- possibly not compilable) and a file containing meta-information about counters. This meta information consists of the correspondences between pairs of counters and the points of interest (simple goals or complex structures such as disjunctions). One additional
pair of counters is added for each predicate; they are the first and the last counters of the predicate. It can be used to evaluate
predicate or procedure coverage. The log file is generated at execution
time, while the other two are created at the same time as the instrumentation.

Currently, the output of this coverage tool is a html file containing the source code in which colours have been added. Only goal coverage is taken into account for the moment -- this can be easily extended using information provided by the log file. Three coverage degrees are distinguished: 
\begin{description}
\item [{covered~(green)}] execution of the goal was successful (at least once) 
\item [{partially~covered~(yellow)}] the goal was executed, but never succeeded 
\item [{not~covered~(red)}] the goal was not executed 
\end{description}
The tool also produces a detailed report which enumerates
all the pairs of counters, and gives the coverage degree of the corresponding
goal. Although it is less visual than the html rendering, this report contains more information, since it also gives the coverage degree of complex structures (disjunctions, switches) and predicates (procedure coverage).


\section{Evaluation}\label{eval}

Table~\ref{tab:Performance-testing} shows the results of a small evaluation of our tool.
Three different properties were examined: for a given testsuite, we measure the size of the generated code, the time needed for its generation by our tool, and the execution overhead of the generated code compared to the execution time of a script that executes the testsuite in an ad-hoc way. The testsuites for the first three procedures were automatically generated by \cite{DBLP:conf/lopstr/DegraveSV08}, while the testsuite for the last procedure (transpose) was a manually written testsuite containing matrices up to a size of 3x3.

\begin{table}[h]
\centering{}\begin{tabular}{|l||>{\centering}p{0.8in}|>{\centering}p{0.4in}|>{\centering}p{0.7in}|>{\centering}p{0.7in}|>{\centering}p{0.4in}|>{\centering}p{0.4in}|}
\hline 
{\scriptsize Goals} & {\scriptsize Determinism} & {\scriptsize Test cases} & {\scriptsize Generated code size } & {\scriptsize Code generation} & \multicolumn{2}{c|}{{\scriptsize Execution (ns)}}\tabularnewline
\cline{6-7} 
 &  &  & {\scriptsize (lines)} & {\scriptsize (ms)} & {\scriptsize gross} & {\scriptsize net}\tabularnewline
\hline
\hline 
{\scriptsize member(in,in)} & {\scriptsize semidet} & {\scriptsize 6} & {\scriptsize 169} & {\scriptsize 12} & {\scriptsize 40} & {\scriptsize 2}\tabularnewline
\hline 
{\scriptsize member(out,in)} & {\scriptsize nondet} & {\scriptsize 4} & {\scriptsize 189} & {\scriptsize 12} & {\scriptsize 40} & {\scriptsize 11}\tabularnewline
\hline 
{\scriptsize bubblesort(in,out)} & {\scriptsize det} & {\scriptsize 24} & {\scriptsize 475} & {\scriptsize 16} & {\scriptsize 60} & {\scriptsize 50}\tabularnewline
\hline 
{\scriptsize transpose(in,out)} & {\scriptsize det} & {\scriptsize 11} & {\scriptsize 288} & {\scriptsize 12} & {\scriptsize 40} & {\scriptsize 8}\tabularnewline
\hline 
\multicolumn{1}{l}{} & \multicolumn{1}{>{\centering}p{0.8in}}{} & \multicolumn{1}{>{\centering}p{0.4in}}{} & \multicolumn{1}{>{\centering}p{0.7in}}{} & \multicolumn{1}{>{\centering}p{0.7in}}{} & \multicolumn{1}{>{\centering}p{0.4in}}{} & \multicolumn{1}{>{\centering}p{0.4in}}{}\tabularnewline
\end{tabular}\caption{\label{tab:Performance-testing}Performance of the testing tool}
\end{table}

As one expects, the size and generation time of the code depends on the number and complexity
of the given set of test cases. Although the execution time also depends on the complexity of the test cases, the evaluation shows a rather constant overhead for the execution of the testcode generated by our framework.


Table~\ref{tab:Performance-coverage} illustrates the performance of the coverage tool. The examined properties
are the size of the instrumented code compared to the size of the original source code, the time needed for instrumentation and the execution overhead caused by the transformation. The table also shows the execution times for both the instrumented and non-instrumented code. The tested procedures are the same as those in Table~\ref{tab:Performance-testing} with the additionnal \texttt{filter\_list}, the latter being a procedure from the code of the test framework itself that does list filtering by a given set of indices. The input parameters are
chosen relatively large in order to produce measurable times (lists of a few hundred to few thousand elements).


%
\begin{table}[h]
\centering{}\begin{tabular}{|l||>{\centering}p{0.5in}|>{\centering}p{0.8in}|>{\centering}p{0.9in}|>{\centering}p{0.5in}|>{\centering}p{0.8in}|}
\hline 
{\scriptsize Goals} & \multicolumn{2}{c|}{{\scriptsize Code size (lines)}} & {\scriptsize Instrumentation } & \multicolumn{2}{c|}{{\scriptsize Execution (ms)}}\tabularnewline
\cline{2-3} \cline{5-6} 
 & {\scriptsize original} & {\scriptsize instrumented} & {\scriptsize (ms)} & {\scriptsize original} & {\scriptsize instrumented}\tabularnewline
\hline
\hline 
{\scriptsize member(in,in)} & {\scriptsize 19} & {\scriptsize 124} & {\scriptsize 32} & {\scriptsize 13} & {\scriptsize 2050}\tabularnewline
\cline{1-1} \cline{5-6} 
{\scriptsize member(out,in)} &  &  &  & {\scriptsize 42} & {\scriptsize 4460}\tabularnewline
\hline 
{\scriptsize bubblesort(in,out)} & {\scriptsize 38} & {\scriptsize 177} & {\scriptsize 53} & {\scriptsize 11} & {\scriptsize 7130}\tabularnewline
\hline 
{\scriptsize transpose(in,out)} & {\scriptsize 58} & {\scriptsize 340} & {\scriptsize 96} & {\scriptsize 3.3} & {\scriptsize 1520}\tabularnewline
\hline 
{\scriptsize filter\_list(in,in,out)} & {\scriptsize 37} & {\scriptsize 191} & {\scriptsize 60} & {\scriptsize 18} & {\scriptsize 4760}\tabularnewline
\hline 
\multicolumn{1}{l}{} & \multicolumn{1}{>{\centering}p{0.5in}}{} & \multicolumn{1}{>{\centering}p{0.8in}}{} & \multicolumn{1}{>{\centering}p{0.9in}}{} & \multicolumn{1}{>{\centering}p{0.5in}}{} & \multicolumn{1}{>{\centering}p{0.8in}}{}\tabularnewline
\end{tabular}\caption{\label{tab:Performance-coverage}Performance of the coverage tool}
\end{table}

Since counter update occurs between goals, the size of the instrumented code (in number of lines) is approximately twice the size of the original code. However, when the switch transformation is applied, additional lines are added
for every switch test statement, but in any case the size of the instrumented code is limited to a few times the size of the original one. However, as can be seen from Table~\ref{tab:Performance-coverage}, the execution time overhead of the instrumented code can be significant. This can be explained in part by the overhead due to the logging operations, in part by the fact that the compiler no longer can perform a number of optimisations. Nevertheless, it should be noted
that the execution overhead is only present when one is measuring test case coverage, and not when one is executing the testsuite.


\section{Conclusion and Future Work}

In this paper, we have presented a test evaluation framework for Mercury. The framework is implemented, and allows one to write and execute a test suite for a Mercury program, and to visualise the code that is (not) covered by the test suite. The implementation is published as open source software \cite{Biener:MercuryTest}. 

Although our implementation is usable for testing small to medium-size Mercury programs, some features are missing in order to make our tool a full-fledged testing tool for Mercury. In the current implementation, the coverage transformation is applied to one module at a time. Consequently, when measuring the coverage of a multi-module program, the generated renaming information files should be merged manually. Even with this workaround, incorrect
results might be produced if two or more examined modules contain predicates having the same name. This is because the lightweight mode analysis that is performed by the tool is not sophisticated enough to choose the correct predicate from
the several candidates in different modules. Neither does the analysis handle partial
instantiation of variables, a somewhat lesser problem given that the Mercury compiler itself has only limited support for these partially instantiated structures.


Also the framework definition itself could be extended. 
Additional conditions could be introduced for the assertions
part of test cases allowing, for example, to state conditions on the relation between the different
solutions of a predicate. One example would be a condition that
checks whether the solutions returned by a call are in ascending order. Further improvements include changing the output format of the framework, for example to xml, in order to make postprocessing of the report easier. A more fundamental improvement would be the introduction of more sophisticated coverage levels into the tool. These improvements are subject to ongoing and further work.



\begin{thebibliography}{10}

\bibitem{Biener:MercuryTest}
Peter Biener.
\newblock {Mercury} test framework.
\newblock http://sourceforge.net/projects/mercurytest, 2010.

\bibitem{DBLP:conf/lopstr/DegraveSV08}
Fran\c{c}ois Degrave, Tom Schrijvers, and Wim Vanhoof.
\newblock Automatic generation of test inputs for {M}ercury.
\newblock In Michael Hanus, editor, {\em LOPSTR}, volume 5438 of {\em Lecture
  Notes in Computer Science}, pages 71--86. Springer, 2008.

\bibitem{55841}
M.~Ducass\'{e} and A.-M. Emde.
\newblock A review of automated debugging systems: knowledge, strategies and
  techniques.
\newblock In {\em ICSE '88: Proceedings of the 10th international conference on
  Software engineering}, pages 162--171, Los Alamitos, CA, USA, 1988. IEEE
  Computer Society Press.

\bibitem{DBLP:conf/haskell/GillR07}
Andy Gill and Colin Runciman.
\newblock Haskell program coverage.
\newblock In Gabriele Keller, editor, {\em Haskell}, pages 1--12. ACM, 2007.

\bibitem{book:glass}
R.~Glass.
\newblock {\em Software runaways: lessons learned from massive software project
  failures}.
\newblock Prentice Hall, 1997.

\bibitem{Haskell:HUnit}
Dean Herington.
\newblock {\em HUnit User's Guide}.
\newblock http://hunit.sourceforge.net/HUnit-1.0/Guide.html, 2002.

\bibitem{12hunt}
A.~Hunt and D.~Thomas.
\newblock Pragmatic unit testing in java with junit.
\newblock Pragmatic Bookshelf, 2003.

\bibitem{book:kaner}
C.~Kaner, J.~Falk, and H.~Q. Nguyen.
\newblock {\em Testing computer software}.
\newblock John Wiley and Sons, 1993.

\bibitem{mycroft:typesprolog}
A.~Mycroft and R.A. O'Keefe.
\newblock A polymorphic type system for {P}rolog.
\newblock {\em Artificial Intelligence}, 23:295--307, 1984.

\bibitem{winko:Myers2004}
Glenford~J. Myers.
\newblock {\em The Art of Software Testing, Second Edition}.
\newblock Wiley, 2004.

\bibitem{Emma:reference_manual}
V.~Roubtsov.
\newblock {\em EMMA: a free Java code coverage tool, Reference Manual}.
\newblock http://emma.sourceforge.net/reference/reference.pdf.

\bibitem{mercury:jlp}
Z.~Somogyi, H.~Henderson, and T.~Conway.
\newblock The execution algorithm of {M}ercury, an efficient purely declarative
  logic programming language.
\newblock {\em Journal of Logic Programming}, 29(4), 1997.

\bibitem{Mercury:reference_manual}
University of Melbourne,
  http://www.cs.mu.oz.au/research/mercury/information/doc-release/reference\_m%
anual.pdf.
\newblock {\em {M}ercury Language Reference Manual}.

\bibitem{SWI:module_test}
Jan Wielemaker.
\newblock {\em Prolog Unit Tests. Manual}.
\newblock http://www.swi-prolog.org/packages/plunit.html, 2006.

\bibitem{zhu:coverage}
H.~Zhu, P.~Hall, and J.~May.
\newblock Software unit test coverage and adequacy.
\newblock {\em ACM Computing Surveys}, 29(4), 1997.

\end{thebibliography}
\end{document}